\documentclass[10pt]{article}
\usepackage{mathptmx}
\usepackage{amssymb,amsmath}
\newtheorem{thm}{Theorem}[section]
\newtheorem{cor}[thm]{Corollary}
\newtheorem{lem}[thm]{Lemma}
\newtheorem{definition}[thm]{Definition}

\def\proof{{\bf Proof. }}
\newtheorem{remark}[thm]{Remark}
\def\be{\begin{equation}}
\def\ee{\end{equation}}
\def\bea{\begin{eqnarray}}
\def\eea{\end{eqnarray}}
\def\bean{\begin{eqnarray*}}
\def\eean{\end{eqnarray*}}
\def\ea{\end{array}}
\def\ds{\displaystyle}

\def\Z{{\mathcal{Z}}}

\def\W{{\mathcal{W}}}
\def\F{{\mathcal{F}}}
\def\P{{\mathcal{P}}}

\def\L{{\mathcal{L}}}
\def\H{{\mathcal{H}}}
\def\S{{\mathcal{S}}}
\def\D{{\mathcal{D}}}

\newcommand{\field}[1]{\mathbb{#1}}
\newcommand{\rz}{\field{R}}
\newcommand{\cz}{\field{C}}
\newcommand{\nz}{\field{N}}

\def\d{{\rm{d}}}
\def\ra{{\rangle}}
\def\la{{\langle}}

\def\fin{{$\hfill\square$}}
\def\hbarr{{\varepsilon}}

\def\Tr{{\rm{Tr}}}
\begin{document}
\title{Mean field limit for bosons and propagation of Wigner measures}
\author{Z.~Ammari\thanks{D\'epartement de Math\'ematiques,
Universit{\'e} de Cergy-Pontoise UMR-CNRS 8088,
2, avenue Adolphe Chauvin
95302 Cergy-Pontoise Cedex France. Email: zied.ammari@u-cergy.fr}
\hspace{.3in} F.~Nier
\thanks{IRMAR, UMR-CNRS 6625, Universit\'e de Rennes I,
campus de Beaulieu, 35042 Rennes
Cedex, France. Email: francis.nier@univ-rennes1.fr} }

\renewcommand{\today}{July 2008}
\maketitle
\begin{abstract}
We consider the N-body Schr\"{o}dinger dynamics of bosons in the mean field limit with
a bounded pair-interaction potential. According to the previous work \cite{AmNi},
the mean field limit is translated into a semiclassical problem with a
small parameter $\varepsilon\to 0$, after introducing an
$\varepsilon$-dependent  bosonic quantization.
The limit is expressed as a push-forward by a nonlinear flow
(e.g.~Hartree) of the associated  Wigner measures. These object and
their basic properties were  introduced in
\cite{AmNi} in the infinite dimensional setting.
The additional result presented here states that the transport by the nonlinear
flow holds for rather general class of quantum states in their mean
field limit.
\end{abstract}

\section{Introduction}
The mathematical analysis
 of the mean field limit of the $N$-body quantum dynamics of bosons
 started with the work of \cite{Hep} and \cite{GiVe}.
Since, the problem has experienced intensive investigations  using
mainly the so-called BBGKY hierarchy method explained in
\cite{Spo}. Interest was focused on studying the cases of singular
interaction potential (see for
example  \cite{BGM}, \cite{EY}, \cite{BEGMY}, \cite{ESY}).

Recently, a new method was given in \cite{FGS} (see also
\cite{FKP}) for a scalar bounded  potential which inspires this work.
The convergence of the quantum dynamics are typically tested
on the above quoted articles, either on coherent states or on Hermite states.
Even when such specific choices are avoided,  the convergence on arbitrary
states still has to be studied.

In the work \cite{AmNi}, Wigner measures  were extended to the
infinite dimensional setting, as Borel probability measures
  under general assumptions. It
was also explained how previous weak formulations of the mean field
limit  are contained in the definition of these asymptotic Wigner measures,
after a reformulation of the $N$-body problem as a semiclassical
problem with the small parameter $\varepsilon=\frac{1}{N}\to 0$. The
basic properties of these Wigner measures were considered and they
were used to check that the mean field dynamics for the coherent
states and Hermite states are essentially equivalent.

\bigskip

In this paper,  the problem of the mean field dynamics is considered
under some restrictive assumptions on the initial data.
The convergence of N-body Schr\"{o}dinger dynamics of bosons in the
mean field limit will be proved for
a  class of density operator sequences, which contains all the common
examples. Remember that contrary to the finite dimensional case no
natural pseudodifferential calculus can be deformed by arbitrary
nonlinear flows, and the propagation of Wigner measures as dual
objects cannot be straightforward in the infinite dimensional case.
The limit is expressed as push-forward by  a nonlinear flow (e.g.~Hartree)
of Wigner measures associated with the sequence of density operators.
The result holds  here when the
pair interaction potential is bounded on
$L^{2}(\mathbb{R}^{2d}_{x,y})$.
This can be considered as a
regular case and subsequent work will be devoted to more singular
cases like in \cite{FKS} with a Coulombic interaction
$V(x-y)=\frac{1}{|x-y|}$ or in the derivation of cubic nonlinear
Schr\"odinger equations with $V(x-y)=\delta(x-y)$ like in the \cite{ESY}.

Since in the literature  the non relativistic and the semi-relativistic
dynamics of bosons were both studied (see \cite{ElSc}), an abstract
setting for the linear part of the flow seems relevant. Examples are
reviewed in the last section.

We keep the same notations as in \cite{AmNi}. The phase-space, a
complex separable Hilbert space, is denoted by $\mathcal{Z}$
with the scalar product $\la .,.\ra$. The symmetric Fock space on $\Z$
is denoted by $\H$ and  $\bigvee^n \mathcal{Z}$ is
the $n$-fold symmetric (Hilbert) tensor product, so that
$\H=\oplus_{n\in \mathbb{N}} \bigvee^n \mathcal{Z}$ as a Hilbert direct sum.
Algebraic direct sums or tensor products are denoted with a $~{alg}~$
superscript. Hence $\H_0=\mathop{\oplus}_{n\in
  \mathbb{N}}^{alg}\bigvee^n \mathcal{Z}$
 denotes the subspace of vectors with a finite number of particles.
For any $p,q\in \mathbb{N}$, the space
$\P_{p,q}(\Z)$ of complex-valued polynomials on $\Z$  is defined with
the following continuity condition: $b\in\P_{p,q}(\Z)$  iff there
exists a unique  $\tilde b\in\L(\bigvee^p\Z,\bigvee^q\Z)$ such that:
$$
b(z)=\la z^{\otimes q}, \tilde{b} z^{\otimes p}\ra\,.
$$
The subspace of $\P_{p,q}(\Z)$ made of polynomials $b$ such that
$\tilde{b}$ is a compact operator  is
denoted by $\mathcal{P}^{\infty}_{p,q}(\Z)$.
The {\it Wick monomial} of  symbol $b\in \P_{p,q}(\Z)$ is the linear
operator
 $b^{Wick}:\H_0\to\H_0$ defined as follows:
\begin{eqnarray*}
b^{Wick}_{|\bigvee^n \Z}=1_{[p,+\infty)}(n)\frac{\sqrt{n!
(n+q-p)!}}{(n-p)!} \;\hbarr^{\frac{p+q}{2}} \;\S_{n-p+q}\left(\tilde{b}\otimes I_{\bigvee^{n-p} \Z}\right)\,,
\end{eqnarray*}
where $\S_{n}$ is the symmetrization
 orthogonal projection from $\otimes^{n}\Z$ onto $\bigvee^n\Z$.
 Remark that $b^{Wick}$ depends on the
scaling parameter $\hbarr$.

Consider a polynomial $Q\in\P_{2,2}(\Z)$ such that $\tilde{Q}\in\L(\bigvee^2\Z)$ is bounded symmetric.
The many-body quantum Hamiltonian of bosons is a self-adjoint operator on  $\H$ having the general shape:
\begin{eqnarray}
\label{hamiltonian}
H_\hbarr=\d\Gamma(A)+
Q^{Wick},
\end{eqnarray}
where $A$ is  a  given self-adjoint operator on $\Z$. The time evolution of the quantum system is given by
$U_{\hbarr}(t)=e^{-i\frac{t}{\hbarr} H_{\hbarr}}$ and $U^0_{\hbarr}(t)=
e^{-i\frac{t}{\hbarr} \d\Gamma(A)}$ for the  free motion. The
commutation $[Q^{Wick}, N]=0$ with the number operator
$N=\d\Gamma(1)=\left(|z|^{2}\right)^{Wick}$, ensures
the essential self-adjointness of $H_{\hbarr}$ on $\D(\d\Gamma(A))\cap\H_0$ and
the fact that both dynamics preserve the number.

Now we turn to the description of  the nonlinear classical dynamics analogues of (\ref{hamiltonian}).\\
Let us first recall some notations from \cite{AmNi}.
Polynomials in $\P_{p,q}(\Z)$ admit Fr\'echet differentials.
For $b\in\P_{p,q}(\Z)$, set
\begin{eqnarray*}
 \partial_{\overline{z}} b(z)[u]=\bar\partial_r b(z+ r u)_{|r=0} ,&& \partial_{z} b(z)[u]=
 \partial_{r} b(z+ r u)_{|r=0}\,,
\end{eqnarray*}
where $\bar\partial_r, \partial_r$ are the usual derivatives over
$\mathbb{C}$. Moreover, $\partial_{z}^{k}b(z)$ naturally belongs to
$(\bigvee^{k}\Z)^{*}$ (i.e.: $k$-linear symmetric functionals) while
$ \partial_{\overline{z}}^{j}b(z)$ is identified via the scalar
product with an element of
$\bigvee^{j}\Z$, for any fixed $z\in \Z$. For $b_{i}\in \P_{p_{i},q_{i}}(\Z)$,
$i=1,2$ and $k\in\mathbb{N}$, set
\begin{equation*}
\partial_z^k b_1 \; .\;\partial_{\bar z}^k b_2 (z) =\la \partial_z^k b_1(z),
\partial_{\bar z}^k b_2(z)\ra_{(\bigvee^k \Z)^{*},\bigvee^{k}\Z}\; \in\P_{p_1+p_2-k, q_1+q_2-k}(\Z)\quad.
\end{equation*}
The multiple {\it Poisson brackets} are defined by
\begin{eqnarray*}
\{b_1,b_2\}^{(k)}=\partial^k_z b_1
.\partial^k_{\bar z} b_2 -\; \partial^k_z b_2 .\partial^k_{\bar z} b_1, &&
\{b_1,b_2\}=\{b_1,b_2\}^{(1)}.
\end{eqnarray*}
The energy functional
\begin{equation*}
    \label{eq.enfunct}
    h(z)=\la z,Az\ra+ Q(z)\,,\;\;\;z\in\D(A),
\end{equation*}
has the  associated vector field $X:\D(A)\to \Z$, $ X(z)=Az+\partial_{\bar z} Q(z)$ and the
nonlinear field equation
\begin{eqnarray*}
\label{hartree}
i \partial_t z_t=X(z_t)
\end{eqnarray*}
with initial condition $z_0=z\in \D(A)$. For our purpose,
 we only need the integral form of the later
equation
\begin{eqnarray}
\label{hartree.int}
 z_t=e^{-i t A} z-i\int_0^t e^{-i (t-s) A} \,\; \partial_{\bar z}Q(z_s) \, ds, \;\mbox{ for } \;z\in\Z.
\end{eqnarray}
The standard fixed point argument implies
 that (\ref{hartree.int}) admits a unique global $C^0$-flow on $\Z$
 which is denoted by
$\mathbf{F}:\rz\times \Z\to \Z$ (i.e.: $\mathbf{F}$ is a $C^0$-map satisfying $\mathbf{F}_{t+s}(z)=
\mathbf{F}_{t}\circ\mathbf{F}_{s}(z)$ and $\mathbf{F}_t(z)$ solves
(\ref{hartree.int}) for any $z\in\Z$).
While considering the evolution of the Wick symbols, the action of the
free flow $e^{-itA}$ will be summarized by the next notation~:
\begin{equation}
  \label{eq.wickfree}
b_{t}=b\circ e^{-itA}~:~\Z\ni z \mapsto b_{t}(z)=b(e^{-itA}z)\,,
\qquad b_{t}\in\oplus^{\rm alg}_{p,q\in\nz} \P_{p,q}(\Z)\,,
\end{equation}
for any $b\in\oplus^{\rm alg}_{p,q\in\nz} \P_{p,q}(\Z)$ and any $t\in \rz$.\\
 Moreover, if $z_t$ solves $(\ref{hartree.int})$,
 and $Q_{t}$ is defined according to (\ref{eq.wickfree}), then
 $w_t=e^{i tA} z_t$ solves the differential equation
$$
\frac{d}{dt}\, w_t=-i \partial_{\bar z} Q_t(w_t)\,.
$$
Therefore for any $b\in\P_{p,q}(\Z)$, the following identity holds
\bean
\frac{d}{dt}\, b(w_t)&=&\partial_{\bar z} b(w_t)[-i \partial_{\bar z}Q_t(w_t)]+
\partial_{z} b(w_t)[-i \partial_{\bar z}Q_t(w_t)]\\&=&i \{Q_t,b\}(w_t).
\eean
This yields for any $z\in \Z$ and $b\in\oplus^{\rm alg}_{p,q\in\nz} \P_{p,q}(\Z)$,
the Duhamel formula
\bea
\label{class-integ-form}
b\circ{\bf F}_t(z)=b_t(z)+i\int_{0}^{t}\left\{Q,b_{t-t_1}\right\}\circ{\bf F}_{t_1}(z)~dt_{1}\,,
\eea
by observing that  $\{Q_{t_1},b\}(w_{t_1})=\{Q,b_{-t_1}\}(z_{t_1})$.

\section{Results}
While introducing or using Wigner measures, all the arguments are
carried out with extracted sequences (or subsequences)
$(\varepsilon_{n})_{n\in \nz}$ such that $\lim_{n\to
  \infty}\varepsilon_{n}=0$, instead of considering a non countable
range $(0, \overline\hbarr)$, $\overline\hbarr>0$, of values for the
small parameter $\varepsilon$.
Without loss of generality (see \cite{AmNi}) one can consider a
countable family $(\rho_{\hbarr_n})_{n\in\nz}$ of density matrices,
$\varrho_{\varepsilon_{n}}\geq 0$, $\Tr\left[\varrho_{\varepsilon_{n}}\right]=1$,
and test them with $\varepsilon_{n}$-quantized (Wick, Weyl or
anti-Wick) observables before taking the limit $\varepsilon_{n}\to 0$.
For the sake of conciseness, the $\varepsilon$ or $\varepsilon_{n}$
parameter does not appear in the notations of quantized observables.

\bigskip

The first condition which characterizes our class of
$\varepsilon_n$-dependent density matrices reads:
\begin{equation*}
   \exists \lambda >0 \;:\; \forall k\in\nz, \;\;{\rm Tr}[ N^k \rho_{\hbarr_n}]\leq \lambda^k  \; \;
   \mbox{ uniformly  in }\, n\in\nz\,, (N=N_{\varepsilon_{n}})\,.   \hspace{.4in} (H0)
\end{equation*}
Wigner measures were constructed in
\cite[Corollary 6.14]{AmNi} for the sequence
$(\rho_{\hbarr_n})_{n\in\nz}$. Possibly
extracting a subsequence still denoted $(\varepsilon_{n})_{n\in\nz}$,
 there exists  a Borel probability measure $\mu$ called
{\it Wigner measure} such that:
\begin{eqnarray}
\label{wigner}
\lim_{\hbarr_{n}\to 0} \Tr[\rho_{\hbarr_{n}} \, b^{Wick}]=\int_{\Z} b(z) \; d\mu(z)\,, \mbox{ for  any  }
\,b\in \oplus_{\alpha,\beta\in\nz}^{\rm
  alg}\P^\infty_{\alpha,\beta}(\Z)\,,
\end{eqnarray}
with again $b^{Wick}=b^{Wick}_{\varepsilon_{n}}$.\\
The statement (\ref{wigner}) does not hold in general for all
 $b\in \oplus_{\alpha,\beta\in\nz}^{\rm alg}\P_{\alpha,\beta}(\Z)$ and
 counterexamples exhibiting the phenomenon of dimensional defect of
 compactness were given in \cite{AmNi}. The extension
 of  (\ref{wigner}) to the larger class  of symbols $\oplus_{\alpha,\beta\in\nz}^{\rm alg}\P_{\alpha,\beta}(\Z)$
 depends on the sequence
 $(\rho_{\hbarr_n})_{n\in\nz}$ and it turns out to be an important fact when studying the mean field limit.
 In the following, a sequence  $(\rho_{\hbarr_{n}})_{n\in\nz}$ with a single Wigner measure $\mu$ will
 have the property $(P)$ when:
\begin{eqnarray*}
\lim_{\hbarr_n\to 0} \Tr[\rho_{\hbarr_{n}} \, b^{Wick}]=\int_{\Z} b(z) \; d\mu(z)\,, \mbox{ for any }
\;b\in \oplus_{\alpha,\beta\in\nz}^{\rm alg}\P_{\alpha,\beta}(\Z)\,.
\hspace{.2in} (P)
\end{eqnarray*}
Here is the main theorem.
\begin{thm}
\label{main-1}
Let the sequence $(\rho_{\hbarr_n})_{n\in\nz}$ of density matrices,
$\varrho_{\varepsilon_{n}}\geq 0$,
$\Tr\left[\varrho_{\varepsilon_{n}}\right]=1$,
$\lim_{n\to\infty}\varepsilon_{n}=0$,  satisfy
 $(H0)$ and $(P)$. Then the limit
\bea
\label{formula-m}
\lim_{n \to \infty} {\rm Tr}[\rho_{\hbarr_{n}} \;\;e^{i\frac{t}{\hbarr_n}
  H_{\hbarr_n}} \; b^{Wick} \;e^{-i\frac{t}{\hbarr_n}
  H_{\hbarr_n}}]=
\int_\Z (b\circ\mathbf{F}_t) (z) \; d\mu\,,
\eea
holds for any $t\in\rz$ and any $\,b\in\oplus_{\alpha,\beta\in\nz}^{\rm
  alg}\P_{\alpha,\beta}(\Z)$ with $b^{Wick}=b^{Wick}_{\varepsilon_{n}}$\,.
\end{thm}
\begin{remark}
\label{trans-measure}
Since $\mathbf{F}$ is a $C^0$-map the r.h.s.~of (\ref{formula-m}) can be written as
\bean
\int_\Z (b\circ\mathbf{F}_t)(z) \; d\mu=\int_\Z b(z) \; d\mu_t,
\eean
where $\mu_t$ is a push-forward measure defined by $\mu_t(B)=\mu(\mathbf{F}_{-t}(B))$, for any Borel set $B$.
\end{remark}
We refer the reader to \cite{AmNi} for the definition of Weyl observables and the Schwartz  class
of cylindrical functions $\S_{cyl}(\Z)$.
\begin{cor}
\label{wigner-measure-id}
Let the sequence $(\rho_{\hbarr_n})_{n\in\nz}$ of density matrices,
$\varrho_{\varepsilon_{n}}\geq 0$,
$\Tr\left[\varrho_{\varepsilon_{n}}\right]=1$,
$\lim_{n\to\infty}\varepsilon_{n}=0$,  satisfy
 $(H0)$ and $(P)$. Then  the limit
\bea
\label{formula-weyl}
\lim_{\varepsilon_n \to 0} {\rm Tr}[\rho_{\hbarr_{n}} \;\;e^{i\frac{t}{\hbarr_n}
  H_{\hbarr_n}} \; b^{Weyl} \;e^{-i\frac{t}{\hbarr_n}
  H_{\hbarr_n}}]=
\int_\Z b\circ\mathbf{F}_t(z) \; d\mu\,,
\eea
holds for any $b\in\S_{cyl}(\Z)$ and any $t\in\rz$.
\end{cor}
\proof
A consequence of Thm.~\ref{main-1} and \cite[Prop.~6.15]{AmNi} is that
the sequence
 $$\rho_{\hbarr_{n}}(t)
=U_{\varepsilon_n}(t) \rho_{\hbarr_{n}} U_{\varepsilon_n}(t)^*$$ admits a single Wigner
measure given by $\mu_t$. Hence, by definition
\bean
\lim_{\varepsilon_n \to 0} {\rm Tr}[\rho_{\hbarr_{n}}(t) \; b^{Weyl}]&=&
\lim_{\varepsilon_n \to 0} \int_{p\Z} \F[b](\xi) \; {\rm Tr}[\rho_{\hbarr_{n}}(t) \; W(\sqrt{2}\pi\xi)] \;L_p(d\xi)\\
&=& \int_{p\Z} \F[b](\xi) \;\int_\Z e^{2\pi i{\rm Re}(z,\xi)} \; d\mu_t(z)\;\;L_p(d\xi)\,.
\eean
\fin \ \\
Another formulation states that the Wigner measure
$\mu_t$ satisfies a transport equation in an integral form.
\begin{cor}
Let $(\rho_{\hbarr_n}(t))_{n\in\nz}$ be as above and let $\mu_t$
denote its Wigner
measure. Then $t\in\mathbb{R}\mapsto\mu_t$ is a solution to
 the transport equation:
\bea
\label{transport.int}
\mu_t(b)=\mu^0_t(b)+ i\int_0^t \mu_s(\{Q,b_{t-s}\})\,ds\,,
\eea
for any $b\in\oplus_{p,q\in\nz}^{\rm  alg}\P(\Z)$ and where $\mu_t^0(B)=\mu(e^{-it A} B)$ for any borel
set $B\,$.
\end{cor}
\proof
The relation
(\ref{transport.int}) is given by testing (\ref{class-integ-form}) on $\mu=\mu_{0}$\,.
\fin
\bigskip

\section{Criteria for the property $(P)$}
In the following, two conditions which ensure the property $(P)$ are
formulated.
 Recall that for any
$P\in {\cal L}(\Z)$ the operator $\Gamma(P)$  acting on  $\H$  is
defined by
$$
\Gamma(P)_{|\bigvee^{n}\Z}=P\otimes P\cdots\otimes P\,
$$
and $\Gamma(P)$ is an orthogonal projector if $P$ is too.
The first criterion is a 'tightness' assumption with respect to the trace
norm of the state
$$
\forall \eta>0, \exists P\in\L(\Z) \mbox{ finite rank orthogonal projector }, \; \forall n\in\nz: \;{\rm Tr}[
(1-\Gamma(P)) \rho_{\hbarr_n}] < \eta
  \; \;\;(T)\,.
$$
The dual version is formulated as  an equicontinuity assumption with respect to the
 Wick symbols:
$$
\forall p,q\in\nz,\, \forall\eta>0, \exists \W_0\subset\L(\bigvee^p\Z,\bigvee^q\Z) \;\; \forall \tilde b\in\W_0,
\forall n\in\nz \;:\;  \left|{\rm Tr}[\rho_{\hbarr_n} b^{Wick}]\right|<\eta \,, \;\;\; (E)
$$
where $\W_0$ is a neighborhood  of zero in $\L(\bigvee^p\Z,\bigvee^q \Z)$ endowed with the $\sigma$-weak topology.
\begin{lem}
Assume that $(\rho_{\hbarr_n})_{n\in\nz}$ satisfies $(H0)$. Then \\
(i) $(T)\Rightarrow (P)$,\\
(ii) $(E)\Rightarrow (P)$.
\end{lem}
\proof
We aim to prove $(P)$ for $b\in\P_{p,q}(\Z)$.\\
(i) Start with
\bean
{\rm Tr}[\rho_{\hbarr_n} \, b^{Wick} ]&=&{\rm Tr}[\rho_{\hbarr_n} \, \Gamma(P)b^{Wick}
\Gamma(P) ]+{\rm Tr}[\rho_{\hbarr_n} \,(1-\Gamma(P)) b^{Wick} \Gamma(P) ]\\
&+&{\rm Tr}[\rho_{\hbarr_n} \, \Gamma(P) b^{Wick} (1-\Gamma(P)) ]+{\rm Tr}[\rho_{\hbarr_n} \, (1-\Gamma(P)) b^{Wick} (1-\Gamma(P)) ]
\eean
Estimate all the terms containing $(1-\Gamma(P))$ in  a similar way. For example, we have
\bea
\left|{\rm Tr}[\rho_{\hbarr_n} \,(1-\Gamma(P)) b^{Wick} \Gamma(P) ]\right|&= &
\left|{\rm Tr}[\la N\ra^{\frac{p+q}{2}} \rho_{\hbarr_n} \,(1-\Gamma(P)) b^{Wick}\la N\ra^{-\frac{p+q}{2}} \Gamma(P) ]\right| \label{b1}\\
&\hspace{-.8in}\leq &  \hspace{-.5in} C_{p,q}(b) \left\|\la N\ra^{\frac{p+q}{2}} \rho_{\hbarr_n}^{1/2}\,\rho_{\hbarr_n}^{1/2}
\,(1-\Gamma(P))\right\|_{1}\label{b2}\\
&\hspace{-.8in}\leq &  \hspace{-.5in}C_{p,q}(b)\left\|\la N\ra^{\frac{p+q}{2}} \rho_{\hbarr_n}\la N\ra^{\frac{p+q}{2}}\right\|_1^{1/2}
\ds\left\| (1-\Gamma(P))\rho_{\hbarr_n}(1-\Gamma(P))\right\|_{1}^{1/2}\label{b3}\\
&\hspace{-.8in}\leq &  \hspace{-.5in} \tilde{C}_{p,q}(b) {\rm Tr}[\rho_{\hbarr_n} (1-\Gamma(P))]^{1/2}\label{b4}.
\eea
First  (\ref{b2}) comes from the number estimate $\left\|b^{Wick}\la
  N\ra^{-\frac{p+q}{2}}\right\|\leq C_{p,q}(b)$
then Cauchy-Schwarz inequality yield (\ref{b3}).  The last estimate
(\ref{b4}) is possible with
$(H0)$. Remark that $\Gamma(P) b^{Wick} \Gamma(P)=\Gamma(P) b(Pz)^{Wick} \Gamma(P)$ and that the polynomial
$b(Pz)\in\P_{p,q}^{\infty}(\Z)$ when $P$ is finite rank orthogonal projector.
The hypothesis $(T)$ and the above argument allow to approximate ${\rm Tr}[\rho_{\hbarr_n} \, b^{Wick} ]$
by the quantity ${\rm Tr}[\rho_{\hbarr_n} \, b(Pz)^{Wick} ]$ using $\eta/3$ argument. \\
Now, write
\bean
\left|{\rm Tr}[\rho_{\hbarr_n} \, b^{Wick} ]-\int_\Z b(z) d\mu\right|&\leq&
\left|{\rm Tr}[\rho_{\hbarr_n} \, \left(b^{Wick} -b(Pz)^{Wick}\right) ]+{\rm Tr}[\rho_{\hbarr_n} \, b(Pz)^{Wick}] -\int_\Z b(Pz) d\mu \right.
\\ & &+\left.\int_\Z [b(Pz)-b(z)] d\mu\right|\,.
\eean
So, the property $(T)$ and $(H0)$ implies $(P)$.\\
(ii) There exists a sequence $b_\kappa\in\P_{p,q}^\infty(\Z)$ such that
$\tilde b_\kappa$ converges in the $\sigma$-weak topology to $\tilde b$. We have
\bea
\label{p-e}
\left|{\rm Tr}[\rho_{\hbarr_n} \, b^{Wick} ]-\int_\Z b(z) d\mu\right|&\leq&
\left|{\rm Tr}[\rho_{\hbarr_n} \, \left(b^{Wick} -b_\kappa^{Wick}\right) ]+\left(
{\rm Tr}[\rho_{\hbarr_n} \, b_\kappa(z)^{Wick}] -\int_\Z b_\kappa(z) d\mu\right) \right.\nonumber
\\ & &+\left.\int_\Z [b_\kappa(z)-b(z)] d\mu\right|\,.
\eea
So, $(P)$ holds by an $\eta/3$ argument and using respectively $(E)$, (\ref{wigner}) and  dominated
convergence  for each term in the (r.h.s.) of (\ref{p-e}).
\fin

\begin{remark}\ \\
1) The space of bounded operators 
$\L(\bigvee^p\Z,\bigvee^q \Z)$ endowed with the $\sigma$-weak topology is not a Baire space when
$\Z$ is infinite dimensional. Otherwise, $(E)$  and hence $(P)$ would
be fulfilled  by any sequence $(\rho_{\hbarr_n})_{n\in \nz}$ satisfying
$(H0)$, according to Banach-Steinhaus Theorem (Uniform Boundedness Principle).\\
2) The hypothesis $(H0)$ in the above lemma, can be replaced by the weaker statement (see \cite[Prop.6.15]{AmNi})
$$
 \exists C>0 : \; \forall k\in\nz,  \; {\rm Tr}[N^{k} \rho_{\hbarr_n}N^{k}]\leq C(Ck)^{k}
$$
uniformly in $\hbarr_n$. This can be interpreted as an
analyticity property of $t\to {\rm
  Tr}[e^{itN^{2}}\varrho_{\varepsilon_{n}}e^{itN^{2}}]$
in $\left\{\left|t\right|<1/C\right\}$, uniformly w.r.t $\varepsilon_{n}$.
\end{remark}

\section{Proof of Thm.~\ref{main-1}}
\begin{definition}
For $m\in \nz$, $r\in\{0,\cdots,m\}$ and
$t_1,\cdots,t_m,t\in\mathbb{R}$, associate with any
$b\in \P_{p,q}(\Z)$ the polynomial:
\begin{eqnarray}
\label{cnr} C^{(m)}_r(t_m,\cdots,t_1,t)=\frac{1}{2^r}\; \sum_{\sharp
\{i:\; \gamma_i=2\}=r}\;  \{Q_{t_m}, \cdots, \{Q_{t_1},b_t
\underbrace{\}^{(\gamma_1)}\cdots\}^{(\gamma_n)}}_{
\gamma_i\in\{1,2\}}
\in\P_{p-r+m,q-r+m}(\Z)\,.
\end{eqnarray}
Note that for shortness the dependence of $C_r^{(m)}(t_m,\cdots,t_1,t)$ on $b$  is not made explicit on the notation and even sometimes we will  write $C_r^{(m)}$. By convention we set $C^{(0)}_0(t)=b_t$.
\end{definition}
\bigskip

We collect some statements from \cite{AmNi}. Remember that $\tilde{b}$
denotes the operator
$\tilde{b}=\frac{\partial_{\bar z}^{q}\partial_{z}^p}{q!p!} b(z)\in
\mathcal{L}(\bigvee^{p}\Z,\bigvee^{p}\Z)$ associated with $b\in \P_{p,q}(\Z)$.
\begin{lem}
\label{tech.lem.1}
Let $b\in\P_{p,q}(\Z)$.\\
(i) The following inequality holds true
\begin{eqnarray*}
\left|\widetilde{\{Q_s,b_t\}^{(2)}}\right|_{\mathcal{L}(\bigvee^{p}\Z,\bigvee^{q}\Z)}
\leq  \; 2[p(p-1)+q(q-1)] \;|\tilde Q| \;|\tilde b|_{\mathcal{L}(
\bigvee^p\Z, \bigvee^q\Z)}\,.
\end{eqnarray*}
(ii) For any $m\in \nz$ and $r\in
\left\{0,1,\ldots,m\right\}$, we have
\begin{eqnarray*}
\left|\widetilde{C^{(m)}_r}\right|_{\mathcal{L}(\bigvee^{p+m-r}\Z,
  \bigvee^{q+m-r}\Z)}\leq
2^{2m-r} \;\ds\left(^m_r\right) \; (p+m-r)^{2r}
\;\frac{(p+m-r-1)!}{(p-1)!} \;|\tilde Q|^m
\;|\tilde b|_{\mathcal{L}(\bigvee^p\Z, \bigvee^q\Z)}\,,
\end{eqnarray*} when $p\geq
q$ with a similar expression when $q\geq p$ (replace $(p+m-r, p-1)$
with  $(q+m-r, q-1)$)\,.
\end{lem}
\proof
See \cite[Lemma 5.8, 5.9]{AmNi}.
\fin
\begin{lem}
\label{tech.lem.2}
For any $\delta>0$ there exists  $ T>0$ such that for all $0<t<T\,$:
\begin{eqnarray}
\label{poisson-brack-conv}
\sum_{m=0}^{\infty}\delta^m \;\int_{0}^{t}d{t_{1}}\cdots\int_{0}^{t_{m-1}}dt_{m}\;\;
\left|\widetilde{C^{(m)}_{0}}(t_{m},\ldots,t_{1},t)\right|_{\L(\bigvee^{p+m}\Z,\bigvee^{q+m}\Z)}\,< \infty
\end{eqnarray}
\end{lem}
\proof
It is enough to bound (\ref{poisson-brack-conv}) in the case $p\geq q$.  Using Lemma \ref{tech.lem.1}
(iii) with $r=0$, we obtain
\bean
\sum_{m=0}^{\infty}\delta^m \;\int_{0}^{t}d{t_{1}}\cdots\int_{0}^{t_{m-1}}dt_{m}\;\;
\left|\widetilde{C^{(m)}_{0}}(t_{m},\ldots,t_{1},t)\right|\leq  2^{p-1} |\tilde b| \;
\sum_{m=0}^{\infty} \left(2^3  \delta \;t\; |\tilde Q|\;\right)^m.
\eean
The r.h.s.~is finite whenever $ \; 0<t<T=(2^3 \;\delta\; |\tilde Q|)^{-1}$.
\fin

\noindent
{\bf Proof of Thm.~\ref{main-1}}\ \\
First consider the following expansion proved in \cite[(50)-(52)]{AmNi} for any positive integer $M$:
\begin{eqnarray*} U_\hbarr (t)^* b^{Wick}U_\hbarr (t)
&=&\sum_{m=0}^{M-1}  i^m  \;\;
 \int_0^t dt_1\cdots\int_0^{t_{m-1}} dt_m \;
 \left[C_0^{(m)}(t_m,\cdots,t_1,t)\right]^{Wick} \\
&&\hspace{-1.7in} + \frac{\hbarr}{2} \sum_{m=1}^{M} i^{m}
\int_0^t dt_1\cdots\int_0^{t_{m-1}} dt_m\; U_\varepsilon(t_m)^*
U^0_\varepsilon(t_m)
\left[\{Q_{t_{m}},C^{(m-1)}_{0}(t_{m-1},\cdots,t_1,t)
\}^{(2)}\right]^{Wick}
U_\varepsilon^0(t_m)^* U_\varepsilon(t_m) \\
&& \hspace{-1.7in} + i^{M} \int_0^t
dt_1\cdots\int_0^{t_{M-1}} dt_M \; U_\varepsilon(t_M)^*
U^0_\varepsilon(t_M)
\left[C^{(M)}_0(t_M,\cdots,t_1,t)\right]^{Wick} U_\varepsilon^0(t_M)^*
U_\varepsilon(t_M)\,,
\end{eqnarray*}
where the equality holds in $\L(\bigvee^s\Z,\bigvee^{s+q-p}\Z)$ for
any $s\in\nz$, $s\geq q-p$. Multiplying on the left the above identity by $\rho_{\hbarr_n}$ and then
using number estimates with the help of $(H0)$, yields an identity on $\L_1(\H)$ on which
we take the trace. This leads to
\bea
\label{p.1}
&&\hspace{-.2in}{\rm Tr}[\rho_{\hbarr_{n}} U_{\hbarr_n} (t)^* b^{Wick}U_{\hbarr_n} (t)]= \sum_{m=0}^{M-1}   i^m
 \int_0^t dt_1\cdots\int_0^{t_{m-1}} dt_m \;
 {\rm Tr}\left[ \rho_{\hbarr_n} \;\left(C_0^{(m)}(t_m,\cdots,t_1,t)\right)^{Wick} \right] \\ &&
 \hspace{-.4in}  +\frac{\hbarr_n}{2} \sum_{m=1}^{M}    i^m \int_0^t dt_1\cdots\int_0^{t_{m-1}} dt_m  \nonumber \\
 \label{p.2}
 && \hspace{.4in}
{\rm Tr}\left[ \rho_{\hbarr_n} U_{\hbarr_n}(t_m)^* U^0_{\hbarr_n}(t_m)\left(\{Q_{t_m}, C_0^{(m-1)}(t_{m-1},\cdots,t_1,t)\}^{(2)}\right)^{Wick}
U^0_{\hbarr_n}(t_m)^* U_{\hbarr_n}(t_m)\right]\\ \label{p.3}&& \hspace{-.4in}
+ i^M  \int_0^t dt_1\cdots\int_0^{t_{M-1}} dt_M \;
 {\rm Tr}\left[ \rho_{\hbarr_n} U_{\hbarr_n}(t_M)^* U^0_{\hbarr_n}(t_M)
 \;\left(C_0^{(M)}(t_M,\cdots,t_1,t)\right)^{Wick} U^0_{\hbarr_n}(t_M)^* U_{\hbarr_n}(t_M)\right].
\eea
The interchange of trace and integrals on the r.h.s.~is justified by the bounds on Lemma \ref{tech.lem.1}.
Lemma \ref{tech.lem.2} implies that the term of (\ref{p.1}) and
(\ref{p.2}) are bounded by
\bean
A_{m}&=& \lambda^{m+\frac{p+q}{2}}\; {\rm sign}(t)^m\int_0^{t} dt_1\cdots\int_0^{t_{m-1}} dt_m \;
  \left|\widetilde{C_0^{(m)}}\right|\\
B_{m}&=& \hbarr_{n}\left|\tilde{Q}\right|
(p+q+m-1)^2 \lambda^{m-1+\frac{p+q}{2}}
\;{\rm sign}(t)^m\int_0^{t} dt_1\cdots\int_0^{t_{m-1}} dt_m \;
\left|\widetilde{C_0^{(m-1)}}\right|\,
\eean
while the remainder (\ref{p.3}) is estimated by
$$
\left|(\ref{p.3})\right|\leq  {\rm sign}(t)^M \int_0^{t} dt_1\cdots\int_0^{t_{M-1}} dt_M \;
\left|\widetilde{C_0^{(M)}}\right|=C_{M}.
$$
By Lemma~\ref{tech.lem.1}, the series $\sum_{m=0}^{\infty}A_{m}$
and $\sum_{m=0}^{\infty}B_{m}$ converge as soon as $|t|<T_0=(2^3 \lambda |\tilde Q|)^{-1}$
while $\lim_{M\to \infty}C_{M}=0$.  Hence the relation
(\ref{p.1})(\ref{p.2})(\ref{p.3}) holds with $M=\infty$ with a
vanishing third term and a second term bounded by $\sum_{m=0}^{\infty}B_{m}=\mathcal{O}(\hbarr_{n})$.
Therefore, we obtain
$$
\lim_{\hbarr_{n}\to 0}{\rm Tr}[\rho_{\hbarr_n} U_{\hbarr_n} (t)^* b^{Wick}U_{\hbarr_n} (t)]-\sum_{m=0}^{\infty}
 i^m \int_0^t dt_1\cdots\int_0^{t_{m-1}} dt_m  {\rm Tr}\left[ \rho_{\hbarr_n}
 \left(C_0^{(m)}(t_m,\cdots,t_1,t)\right)^{Wick} \right]=0.
$$
Owing to the condition $(P)$ which provides the pointwise convergence
and the uniform bound of $\sum_{m=0}^{\infty}A_{m}$, the Lebesgue's convergence theorem implies
\bea
\label{convergence}
\lim_{\hbarr_n\to 0}\; \sum_{m=0}^{\infty}   i^m
 \int_0^t dt_1\cdots\int_0^{t_{m-1}} dt_m \; && \hspace{-.2in} {\rm Tr}\left[ \rho_{\hbarr_n} \;\left(C_0^{(m)}(t_m,\cdots,t_1,t)\right)^{Wick} \right]
 = \nonumber \\
  &&\sum_{m=0}^{\infty}   i^m \int_0^t dt_1\cdots\int_0^{t_{m-1}} dt_m \;\int_{\Z} C_0^{(m)}(t_m,\cdots,t_1,t;z) \;d\mu\,.
\eea
Now, we interchange the sum over $m$ and the integrals on $(t_1,\cdots,t_m,t)$
with the integral over $\Z$ on (\ref{convergence}) simply with a
Fubini argument  based on the absolute convergence (written here for $t>0$):
\bean
\sum_{m=0}^{\infty}
 \int_0^t dt_1\cdots\int_0^{t_{m-1}} dt_m \;
 &&\hspace{-.2in} \int_\Z \,\left| C_0^{(m)}(t_m,\cdots,t_1,t;z)\right| \;d\mu \leq \\
 &&\sum_{m=0}^{\infty} \;\left(\int_\Z |z|^{p+q+2m}  \;d\mu \right)\;
 \int_0^t dt_1\cdots\int_0^{t_{m-1}}dt_m  \left|\widetilde{C_0^{(m)}}(t_m,\cdots,t_1,t)\right| \;  .
\eean
Again $(H0)$ and $(P)$ imply that for all $k\in\nz$ there exists $\lambda >0$ such that
$$
\ds\int_\Z \;|z|^{2k}\;d\mu=\lim_{\hbarr_n \to 0} {\rm Tr}[\rho_{\hbarr_n} \;(|z|^{2k})^{Wick}]
=\lim_{\hbarr_n \to 0} {\rm Tr}[\rho_{\hbarr_n} N^{k}] \leq \lambda^k.
$$
Hence, Lemma \ref{tech.lem.2} yields for $|t|<T_0$:
\bean
\lim_{\hbarr_n\to 0}{\rm Tr}[\rho_{\hbarr_n} U_{\hbarr_n} (t)^* b^{Wick}U_{\hbarr_n} (t)]&=&
\sum_{m=0}^{\infty}  i^m  \;\;
 \int_0^t dt_1\cdots\int_0^{t_{m-1}} dt_m \;\int_{\Z} C_0^{(m)}(t_m,\cdots,t_1,t;z) d\mu\\
 &=&\int_{\Z} \;\sum_{m=0}^{\infty}   i^m \int_0^t dt_1\cdots\int_0^{t_{m-1}} dt_m \;
C_0^{(m)}(t_m,\cdots,t_1,t;z) \;d\mu\,,
\eean
where the integrand $\ds\sum_{m=0}^{\infty}   i^m \int_0^t dt_1\cdots\int_0^{t_{m-1}} dt_m \;
C_0^{(m)}(z)$  is a convergent series in $L^1(\mu)$.\\
The last step is the identification of the limit with the r.h.s.~of
(\ref{formula-m}). Indeed,  an iteration of (\ref{class-integ-form})
 reads
\begin{eqnarray*}
b(z_{t})
=b_t(z)+  i \int_0^t  \; \{Q_{t_1},b_t\}(z)\;dt_1+ i^2 \int_0^t
dt_1\int_0^{t_{1}} dt_2 \; \{Q_{t_2},\{Q_{t_1},b_t\}\}(e^{i t_2 A} z_{t_2})\,,
\end{eqnarray*}
 after setting $z_{t}=\mathbf{F}_{t}(z)$ and defining the Wick
symbols $b_{t}$ and $Q_{t}$ according to \eqref{eq.wickfree}.
By induction we obtain for any $M>1$:
\begin{eqnarray*}
b\circ{\bf F}_t(z)
&=&b_t(z)+\sum_{m=1}^{M-1}  i^m  \;\;
 \int_0^t dt_1\cdots\int_0^{t_{m-1}} dt_m \; \;C_0^{(m)}(t_m,\cdots,t_1,t;z) \label{idf.1}\\
&+&  i^{M} \int_0^t
dt_1\cdots\int_0^{t_{M-1}} dt_M \; \;C^{(M)}_0(t_M,\cdots,t_1,t;e^{i t_M A} z_{t_M}) \,.\label{idf.2}
\end{eqnarray*}
An integration with respect to the measure $\mu$ leads to
\bean
\int_{\Z} \,b\circ{\bf F}_t(z) \;d\mu&=& \sum_{n=0}^{M-1}  i^n  \;\;
 \int_0^t dt_1\cdots\int_0^{t_{n-1}} dt_n \;\int_{\Z} C_0^{(n)}(t_n,\cdots,t_1,t;z) d\mu\\
&+&  i^{M} \int_0^t
dt_1\cdots\int_0^{t_{M-1}} dt_M \; \int_{\Z} C^{(M)}_0(t_M,\cdots,t_1,t;e^{i t_M A} z_{t_M})  \;d\mu\,.
\eean
Again the uniform estimate $\sum_{m=0}^{\infty}A_{m}$ when $|t|<T_{0}$
and  $\lim_{M\to  \infty}C_{M}=0$,
 allow to take the limit as $M\to\infty$. This implies
for $|t|<T_0$
$$
\int_{\Z} \,b\circ{\bf F}_t(z) \;d\mu= \sum_{m=0}^{\infty}  i^m  \;\;
 \int_0^t dt_1\cdots\int_0^{t_{m-1}} dt_m \;\int_{\Z} C_0^{(m)}(t_m,\cdots,t_1,t;z) \;d\mu.
$$
This proves the result for $|t|<T_0$ and it is extended to any time by
the next iteration argument.
Indeed,  it is clear  that $\rho_{\hbarr_{n}}(t)=U_{\hbarr_n} (t)\rho_{\hbarr_{n}} U_{\hbarr_n} (t)^*$
satisfies $(H0)$ since $U_{\hbarr_n} (t)$ commute with $N$.
The property  $(P)$ holds  for $\rho_{\hbarr_{n}}(t)$ when $|t|<T_0$ by Remark \ref{trans-measure} and
Corollary \ref{wigner-measure-id}.  For $t,s$ such  that
$|t|,|s|<T_0$, the sequence $(\rho_{\hbarr_{n}}(t))_{n\in\nz}$ satisfies
$(H0)$ and $(P)$. Therefore, the result for short times yields
\bean
\lim_{\hbarr_n\to 0}{\rm Tr}[\rho_{\hbarr_n}(t) U_{\hbarr_n} (s)^* b^{Wick}U_{\hbarr_n} (s)]
&=& \int_{\Z}  b\circ{\bf F}_s(z)\;d\mu_t\,=\int_{\Z}  b\circ{\bf F}_{t+s}(z)\;d\mu.
\eean
\fin

\begin{remark} As by product we have for any $b\in  \oplus_{\alpha,\beta\in\nz}^{\rm alg}\P_{\alpha,\beta}(\Z)$
\bea
\label{conv-lmu}
b\circ{\bf F}_t(z)=L^1(\mu)-\sum_{m=0}^{\infty}   i^m \int_0^t dt_1\cdots\int_0^{t_{m-1}} dt_m \;
C_0^{(m)}(t_m,\cdots,t_1,t;z)\,.
\eea
Moreover, the arguments used in the proof of Thm.~\ref{main-1} can not ensure the pointwise absolute
convergence of the r.h.s. (\ref{conv-lmu}) for all $z\in\Z$.
\end{remark}

\section{Examples}
\underline{\sf \bf Models:}

\noindent
M1) Let $\Z=L^2(\rz^d,dx)$, $A=D_x^2+U(x)$ self-adjoint  and $Q$ is a multiplication operator by
$\frac{1}{2} V(x-y)$ with $V\in L^\infty(\rz^d)$.\\
M2)  Let $\Z=L^2(\rz^d,dx)$, $A=\sqrt{D_x^2+m^2}+U(x)$ self-adjoint
and $Q$ as above.\\
M3) When $\Z=\cz^{d}\sim \rz^{2d}_{x,\xi}$, one recovers the standard
semiclassical limit problem and the condition $(P)$ is always
satisfied if $(H0)$ is satisfied. We refer for example the reader to
\cite{CRR} \cite{Ger} \cite{GMMP} \cite{HMR} \cite{LiPa} \cite{Mar} \cite{Rob}
for various results about this topic.

\bigskip
\noindent
\underline{\sf\bf Density operator Sequences:}

\noindent
1) Every sequence $(\rho_{\hbarr_n})_{n\in\nz}$ valued in a compact set of the Banach space of trace class operators has the
Wigner measure $\delta_0$. If in addition $(\rho_{\hbarr_n})_{n\in\nz}$ satisfies $(H0)$ then $(P)$ holds true. \\
2) Let $(\rho_{\hbarr_n})_{n\in\nz}$ as in 1) and satisfying $(H0)$ and let
$(z_n)_{n\in\nz}$ be a sequence of $\Z$ such that $\lim_{n\to \infty}$ $\left|z_{n}-z\right|=0$.
Then
$\tilde\rho_{\hbarr_n}=W(\frac{\sqrt{2}}{{i\hbarr}}z_{n})
\rho_{\hbarr_n} W(-\frac{\sqrt{2}}{{i\hbarr}}z_{n})$
admits the unique
Wigner measure
$\mu=\delta_{z}$ where $z$ and (P) holds true. The push-forward measure is
$\mu_t=\delta_{z_t}$. \\
3) Let $(z_n)_{n\in\nz}$ be a sequence valued in a compact set of $\Z$. So $\rho_{\hbarr_n}=|z_n^{\otimes [\hbarr_n^{-1}]}\ra \la
z_n^{\otimes [\hbarr_n^{-1}]}| $  satisfies $(H0)$ and the property $(P)$ and admits the Wigner measures
$\frac{1}{2\pi} \int_0^{\pi} \delta_{e^{i\theta}z} d\theta$ where $z$ is any cluster point of $(z_n)_{n\in\nz}$.
Several other examples can be obtained by superposition, see \cite{AmNi}.\\
4) Let $(z_n)_{n\in\nz}$ be a sequence such that $|z_n|=1$ in $\Z$ converging weakly to $0$. Then  $(P)$ fails for
$\rho_{\hbarr_n}= |E(z_{n})\ra\la E(z_{n})|$ with
$E(z_{n})=W(\frac{\sqrt{2}}{{i\hbarr}}z_n)|\Omega\ra $,
although $(H0)$ holds.

\medskip
\bibliographystyle{empty}

\end{document}